\documentclass[12pt
,notitlepage
,superscriptaddress
]{revtex4}
\usepackage{graphicx}
\usepackage{amsmath}
\usepackage{comment}

\begin{document}

\title{Unravelling Nanoconfined Films of Ionic Liquids}

\author{Alpha A Lee }
\affiliation{Mathematical Institute, Andrew Wiles Building, University of Oxford, Woodstock Road, Oxford OX2 6GG, United Kingdom}

\author{Dominic Vella}
\affiliation{Mathematical Institute, University of Oxford, Woodstock Road, Oxford OX2 6GG, United Kingdom}


\author{Susan Perkin}
\affiliation{Department of Chemistry, University of Oxford, Oxford, OX1 3QZ, United Kingdom}

\author{Alain Goriely}
\affiliation{Mathematical Institute, University of Oxford, Woodstock Road, Oxford OX2 6GG, United Kingdom}

\begin{abstract}
The confinement of an ionic liquid between charged solid surfaces is treated using an exactly solvable 1D Coulomb gas model. The theory highlights the importance of two dimensionless parameters: the fugacity of the ionic liquid, and the electrostatic interaction energy of ions at closest approach, in determining how the disjoining pressure exerted on the walls depends on the geometrical confinement. Our theory reveals that thermodynamic fluctuations play a vital role in the ``squeezing out'' of charged layers as the confinement is increased. The model shows good qualitative agreement with previous experimental data, with all parameters independently estimated without fitting.      
\end{abstract}

\makeatother
\maketitle

\section{Introduction}
Room temperature ionic liquids are used in diverse fields ranging from solvents in chemical synthesis \cite{welton1999room,hallett2011room}, to electrochemical supercapacitors \cite{brandt2013ionic} and lubricants \cite{smith2013quantized}. The non-volatility and wide electrochemical window of stability of ionic liquids render them good candidate electrolytes for electrochemical and energy storage applications, especially in nanodevices where a large voltage is applied over small length-scales (for a recent review of ionic liquids at electrified interfaces, see \cite{fedorov2014ionic}). In nanotechnological applications, such as nanoporous supercapacitors and lubricants for microelectromechanical devices, ionic liquids are under severe geometric confinement between charged surfaces. 

Previous studies focused on the effect of strong electrostatic correlations on the semi-infinite electrode-electrolyte interface. Experiments (e.g \cite{alam2007measurements,alam2007measurementsJPC,alam2008capacitance,islam2008electrical,lockett2008differential}), theories (e.g \cite{eigen1954thermodynamics, borukhov1997steric,borukhov2000adsorption}) and simulations (e.g \cite{esnouf1988computer, lamperski2008grand,trulsson2010differential,vatamanu2010molecular,fedorov2010double,Hu2013}, or a recent review \cite{merlet2013computer} ) have shed light on the structure of bulk ionic liquids near charged surfaces. Those pioneering works revealed the effect of ion size in determining the interfacial capacitance, and established that there is a region of closely packed ions near a highly charged surface, followed by a gradual (potentially oscillatory) decay of ion density into the bulk. 

Analysis of the long-distance asymptotics of the direct correlation function obtained by the Ornstein-Zernicke equation reveals \cite{attard1993asymptotic, leote1994decay, attard1996electrolytes} that the density-density and charge-charge correlation functions decay exponentially with or without an oscillatory component. Increasing the ion density (or chemical potential) at fixed temperature causes a crossover from exponential decay of both charge-charge and density-density correlation functions to oscillatory decay of the charge-charge correlation function. A further increase in density triggers the density-density correlations to decay in an oscillatory manner. 

The oscillatory decay of charge-charge correlations can be rationalised as the result of a competition between the long-ranged ion-ion Coulomb interactions and the steric constraint of packing counterions around the central ion. This causes the ionic atmosphere near the ion to overcompensate the bare ion charge. Crossover from monotonic to oscillatory decay of the density-density correlation function is induced by increased steric correlation as the ion density increases, and is also reported in simple square-well fluids \cite{fisher1969decay,evans1993asymptotic}. This is known in the context of single-component fluids as the Fisher-Widom transition. Experimentally X-ray reflectivity \cite{mezger2008molecular} and AFM \cite{hayes2009pronounced} measurements showed an oscillatory decay of charge density away from a highly charged interface, which is corroborated by a recent analytical solution of a 1D lattice Coulomb gas model of ionic fluids \cite{demery2012overscreening,demery2012one}. 

The behaviour and structure of severely confined ionic systems are more intricate. Recent surface force balance experiments by Perkin $et \; al.$ \cite{perkin2010layering,perkin2011self} demonstrated that the disjoining pressure across a nanometre-thick ionic liquid film confined between two (negatively charged) atomically flat mica surfaces increases in an oscillatory manner as the separation between surfaces decreases (see Figure 1 for schematic representation of these experiments and the typical results). The peaks in the disjoining pressure were ascribed qualitatively to squeezing out of layers of ions close to the interface with the position of the peak being an indicator of the ion radii. The position of first peak in those experiments corresponds roughy to the width of the ions, as would be expected on steric grounds. However, the subsequent peaks are much broader and are located further away from integer multiples of the ion radii. This suggests that one must move beyond a simple qualitative geometric description to understand the behaviour of a confined ionic liquid. In particular, how thermodynamic fluctuations and strong electrostatic correlations affect the disjoining pressure in systems under severe geometric confinement has not been elucidated theoretically. Here, we aim to fill this gap. 

\begin{figure}
\centering
\includegraphics[scale=0.5]{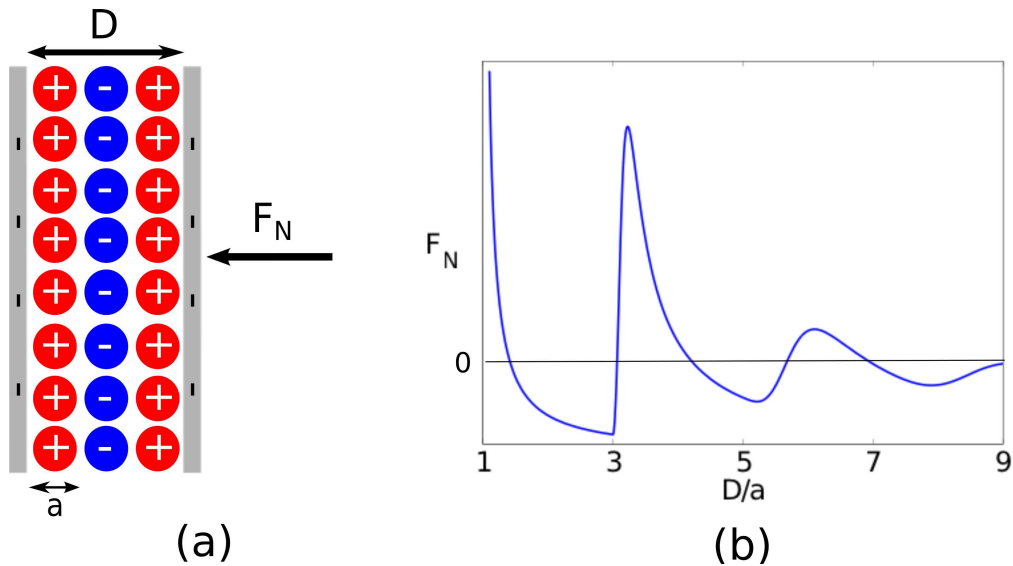}
\caption{Cartoon of the surface force balance experiments performed by Perkin $et \; al.$ \cite{perkin2010layering,perkin2011self}: Panel (a) shows the layering structure of an ionic liquid confined between charged surfaces as well as the force, $F_N$, required to impose a given plate separation $D$; (b) shows the typical dependence of the normal force on the surface separation. }
\end{figure}

Although the pioneering work on semi-infinite electrode-electrolyte interface forms a basis for the general understanding of strongly correlated Coulomb systems, we stress that the oscillatory disjoining force observed in a geometrically confined ionic liquid has a very different physical origin, and should not be conflated with the oscillatory density-density and charge-charge correlation functions observed in the semi-infinite case. Rather, molecular layering of ions close to the charged surface, and the squeezing out of these molecular layers from the slit as the surface separation decreases are the crucial aspects of this intrinsically nanoscale phenomenon. 

We develop a novel and exactly solvable theory of the disjoining pressure in a nanoconfined ionic liquid by considering the system as a 1D Coulomb gas with hard core repulsion. In experimentally relevant systems, the separation between the charged surfaces ($O(\mathrm{nm})$) is much smaller then the typical lateral lengthscale ($O(\mathrm{cm})$). The highly charged surface promotes ordering of anions and cations into slabs, and motivates treating the system as an 1D collection of charged slabs rather than a 3D system.

Our model reproduces the full range of experimentally observed physical phenomenology, and unravels the key roles that the bulk chemical potential and strength of electrostatic interaction play in determining the disjoining pressure.  

\section{The Model} 
\begin{figure}
\includegraphics[scale=0.5]{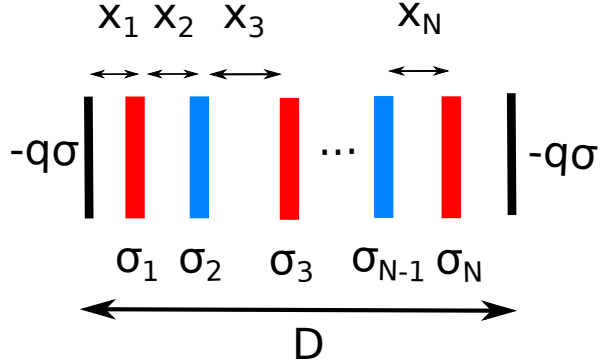}
\caption{Schematic of the 1D Coulomb system. Hard slabs of charge $\pm \sigma$ (cations and anions denoted in this schematic by red and blue respectively) are confined within a slit width $D$. The charge on each of the walls that make up the slit is $-q \sigma$}
\label{schematic}
\end{figure}

We consider a 1D system of hard slabs with width $a$ (physically corresponding to the ion diameter) and fixed charges $\pm \sigma$.  This system is confined between rigid charged surfaces with surface charge density $-q \sigma$ each and coupled to a bath of bulk ionic liquid. To ensure overall electroneutrality, the sum of the total charge of the slabs must equal $-2q\sigma$ (see Figure \ref{schematic}). As such $q$ must be a half integer or integer --- this technical restriction is due to neglecting fluctuations in the surface charge density. 

We note that the thermodynamic properties of a 1D Coulomb gas without hard-core exclusion have been studied extensively in \cite{lenard1961exact, edwards1962exact}. Excluded volume effects have been accounted for by restricting ions to lie on a lattice \cite{demery2012overscreening,demery2012one}. This lattice Coulomb gas model, whilst revealing important qualitative insights for the arrangement of ions, presents artefacts due to the discrete nature of the lattice. This is particularly prevalent in the strongly confined limit, where there are only a few lattice sites and treating the positions of each ion as a discrete variable has the rather unphysical consequence that the disjoining pressure is only defined for discrete surface separations.  Here, we remove this restriction, considering instead a Coulomb gas in which the position of each slab is a continuous variable subject to a hard-core exclusion. \footnote{Effectively 1D electrostatics also emerges in systems with large dielectric contrast, see \cite{kamenev2006transport, zhang2006ion}}

In 1D, the dimensionless electrostatic interaction energy, $v_{ij}(r)$, between 2 slabs separated by a distance $r$ takes the form 
\begin{equation}
\beta v_{ij}(r) = \begin{cases} 
- \Xi S_i S_j |r|  & |r| \ge 1 \\ 
\infty & |r|<1, 
\end{cases}  
\end{equation}
(see Ref \cite{demery2012one}) where $\beta = 1/(k_B T)$, $S_{i}$ and $S_{j}$ can take values $\pm 1$, the lengthscales are non-dimensionalised with respect to the ion diameter $a$, and the electrostatic parameter
\begin{equation}
\Xi = \beta \frac{e^2 \sigma^2 A a}{4\pi \epsilon_0 \epsilon}
\label{eq_for_xi}
\end{equation}
is the ratio of electrostatic interaction energy at closest approach relative to the thermal energy $k_B T$, with $A$ the area of the slab, $\epsilon$ the dielectric constant of the medium, $\epsilon_0$ the permittivity of free space, and $e$ the fundamental charge. 

We note that the 1D model assumes a regular and segregated arrangement of cations and anions in the direction perpendicular to the boundary. This is expected to hold when the surface is strongly charged and thus the ions closest to the surface form a densely packed cation layer, followed by a densely packed anion layer which is attracted to the cation layer, and so on. In addition, the assumption that the slabs have fixed charge relies on small surface separation, and thus the layers are still purely cations or anions. 

The canonical partition function of $m$ cations and $n$ anions read
\begin{equation}
Q_{m,n} = \frac{1}{m! n!} \sum_{\{ \sigma \}} \delta_{m-n, 2q} \int_0^D \mathrm{d}r_1 ... \int_0^D \mathrm{d}r_N \; e^{-\beta \sum_{i<j} v_{ij}(r_i-r_j)}, 
\end{equation} 
where the Kronecker delta function imposes global electroneutrality, $ \sum_{\{ \sigma \}} $ denotes summing over all possible charge configurations and $r_i \in [0,L]$ denotes the position of the charges with $r_0$ and $r_{N+1}$ being the positions of the charged surfaces which are fixed, and $L$ being the separation between the charged surfaces. The integral can be rewritten taking advantage of the fact that the slabs cannot overlap. By denoting the separation between the $i^{\mathrm{th}}$ and $(i+1)^{th}$ slab by $x_{i}$ (see Figure \ref{schematic}), we have 
\begin{equation}
\int_0^D \mathrm{d}r_1 ... \int_0^D \mathrm{d}r_N = \int_{1/2}^{D- (m+n) + 1/2} \mathrm{d}x_1 ...  \int_{1}^{D- \sum_{i=1}^{j-1} x_i  - (m+n)+  j/2}  \mathrm{d}x_j ... \int_{1}^{D- \sum_{i=1}^{N- 1} x_i -1/2} \mathrm{d}x_{N}. 
\label{int_arrangement}
\end{equation}

In experimental configurations, the confined ionic liquid is in thermal equilibrium with the surrounding bulk fluid. As such, we transform the partition function in the Canonical Ensemble into the Grand Canonical ensemble to take into account fluctuations in the number of slabs --- the fluctuations correspond physically to squeezing out of ion layers. Introducing the fugacity $\lambda$, which is assumed to the the same for cations and anions, the Grand Canonical partition function reads
\begin{equation}
\Lambda = \sum_{m,n=0}^{\left\lfloor{\frac{(m+n)a}{L}}\right\rfloor} \lambda^{m+n} Q_{m,n}. 
\end{equation} 
The fugacity is a measure of the bulk cohesive energy --- the larger the fugacity is, the more dense is the bulk ionic fluid. 

The key physical quantity of practical interest is the disjoining pressure. This can be computed from the Grand Canonical partition function by noting that the free energy takes the form $F = - \beta^{-1} \log \Lambda$, and that the disjoining pressure is
\begin{equation} 
P = -\frac{1}{A} \frac{\partial F}{\partial D}. 
\end{equation}

Before illustrating the typical results of the model, we briefly discuss typical parameter values for ionic liquids in a surface force balance experiment. The electrostatic parameter $\Xi$ depends on the pairwise interaction between ions located in adjacent slabs, and thus the lateral packing geometry in the slabs. The fact that viscous liquid-like behaviour is still observed for nanoconfined ionic liquid \cite{bou2010nanoconfined} suggests that the electrostatic interaction is comparable to thermal energy, thus $\Xi$ is expected to be order unity or less, and can be estimated using tribological measurements, as discussed later. The fugacity, on the other hand, is a bulk property of the ionic liquid, and is related to bulk density and affinity of the ionic liquid to the slit. As such, we will treat it as an effective parameter and provide estimates for it via interfacial tension in the section below. 

\section{Like-Charged Interfaces} 
In surface force balance experiments, the film of ionic liquid is usually confined between atomically flat and negatively charged mica surfaces (e.g. \cite{perkin2011self,smith2013quantized}). The normal force measured in these experiments is oscillatory, and the period of the oscillation suggests that an odd number of layers is confined between surfaces. As such, we set $q = 1/2$, $i.e.$ the ions overcompensate the surface charge of the interface. 

Qualitative behaviour of the dependence of the pressure on the surface separation can be seen in the narrow separation limit. The partition function for narrow separations ($1<D<3$), for which only one layer is allowed is
\begin{equation}
Q_1 = \int_{1/2}^{D - 1/2} \; \mathrm{d}x_1 \; e^{- \frac{\Xi D}{4} } = e^{-\frac{\Xi}{4} D} (D - 1). 
\end{equation}
Thus for this one slab system, the pressure is simply
\begin{equation}
\frac{P}{P_0} = \frac{1}{D-1} - \frac{\Xi}{4}, 
\label{p_diss_1}
\end{equation}
with $P_0 = k_B T/a A$ the pressure scale. The first term in Equation (\ref{p_diss_1}) originates from the entropy of confinement, while the second term is due to electrostatic interactions between the slabs. Equation (\ref{p_diss_1}) shows that $P>0$ for small separations as the loss of translational entropy penalises confinement. However, for $\Xi>2$, the disjoining pressure is negative for intermediate separations when the effect of electrostatic correlations dominate the effect of confinement. An expression for the disjoining pressure similar to Equation (\ref{p_diss_1}) has been obtained via a systematic perturbative expansion of the full 3D, counterion only Coulomb gas in the limit of highly charged surfaces (strong-coupling electrostatics) \cite{netz2001electrostatistics,naji2013perspective}. There, the partition function is dominated by single-particle contributions from the interaction between counterions and charged surfaces. 

\begin{figure}
\centering
\includegraphics[scale=0.7]{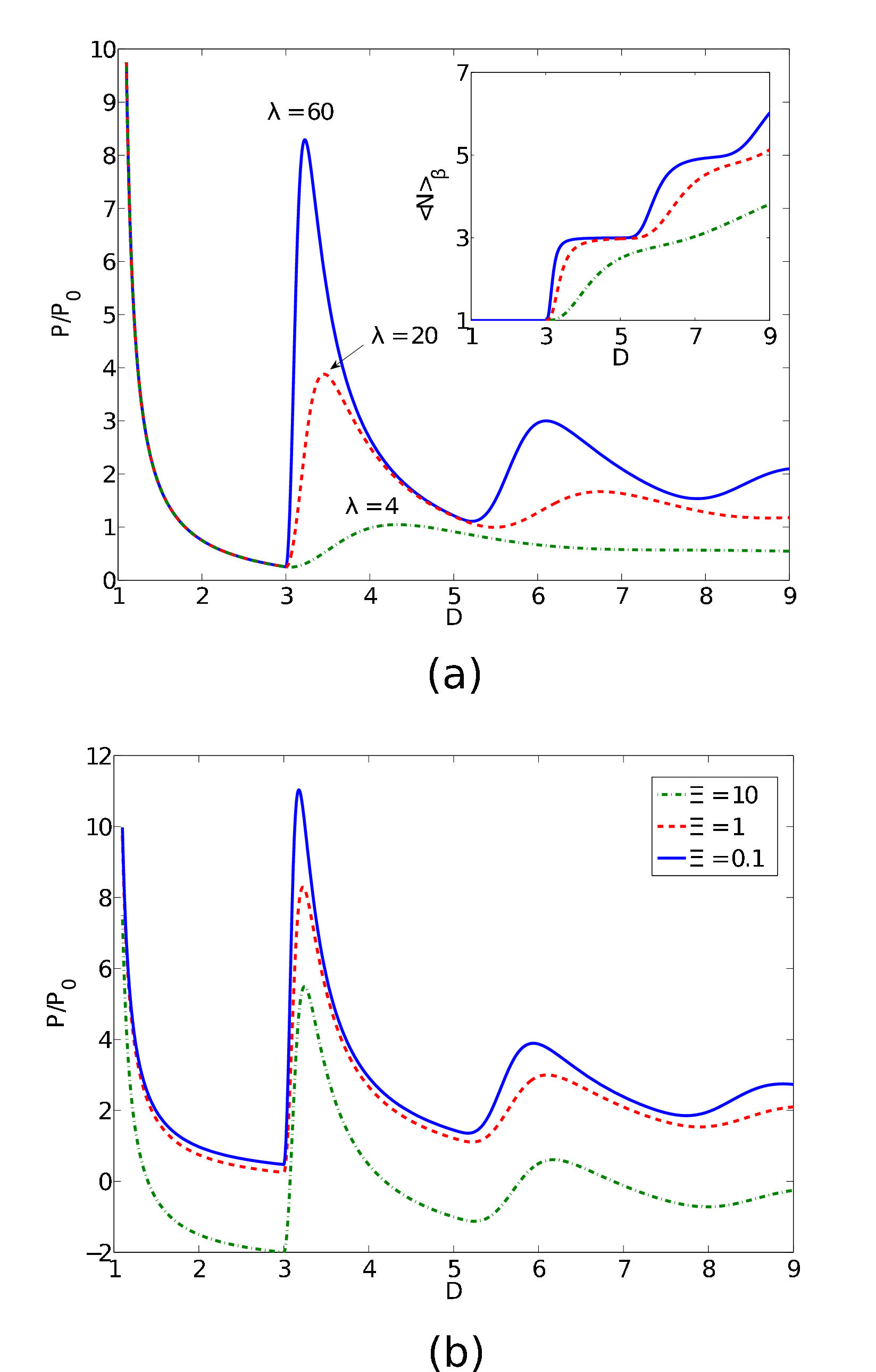}
\caption{The disjoining pressure shows abrupt peaks as the separation increases for large fugacity when ions are confined between like-charged surfaces. (a) The main panel shows the disjoining pressure plotted as a function of separation, and the inset shows the occupancy plotted as the function of separation. $P_0 = k_B T/a A$ is the pressure scale. (b) Decreasing the electrostatic parameter $\Xi$ promotes thermal fluctuations and increases the peak pressure. The plot shows disjoining pressure as a function of separation plotted for different values of the electrostatic parameter $\Xi$ for fugacity $\lambda=60$.}
\label{pressure_like_charge}
\end{figure}

For larger slit widths, multiple layers of ions could be found in the slit. The partition function for the multiple layers of ions can be computed analytically by evaluating the integrals in Equation (\ref{int_arrangement}) via Mathematica. The resulting expressions are rather cumbersome and as such we do not include them here. Figure \ref{pressure_like_charge}(a) shows a peak in disjoining pressure as two new layers enter the slit (we note that global electroneutrality requires layers to enter in positive-negative pairs). Note that the magnitude of the peaks decrease as the separation increases, reflecting the decreased thermodynamic driving force for adsorption when there are multiple layers of ions already present in the slit. The peaks are more pronounced for large fugacity as the ionic fluid favours a densely packed configuration, and thus would fill the slit as soon as the separation exceeds the minimum width at which it is geometrically possible to do so. However, for lower fugacities, the peaks are broader and shift away from the minimum separation dictated by geometry.  As the separation increases they become less ``periodic''. This suggests that directly relating the peaks in disjoining pressure measured from surface force balance experiments to ionic radii is too simplistic --- thermodynamic fluctuations play an important role in the position of those peaks. 

The effect of varying the electrostatic parameter $\Xi$ is shown in Figure \ref{pressure_like_charge}(b). The peak in disjoining pressure becomes more pronounced as $\Xi$ decreases, and for large $\Xi$ the pressure becomes negative, reflecting an attractive force. Physically, this attraction occurs because in the ground state configuration of alternating positive and negative charges, which is thermodynamically favourable for large $\Xi$, the Hamiltonian is identically $H=\Xi D/4$ regardless of the number of slabs. Thus thermal fluctuations are the sole driving force for the insertion of addition layers. 

\begin{figure}
\centering
\includegraphics[scale=0.4]{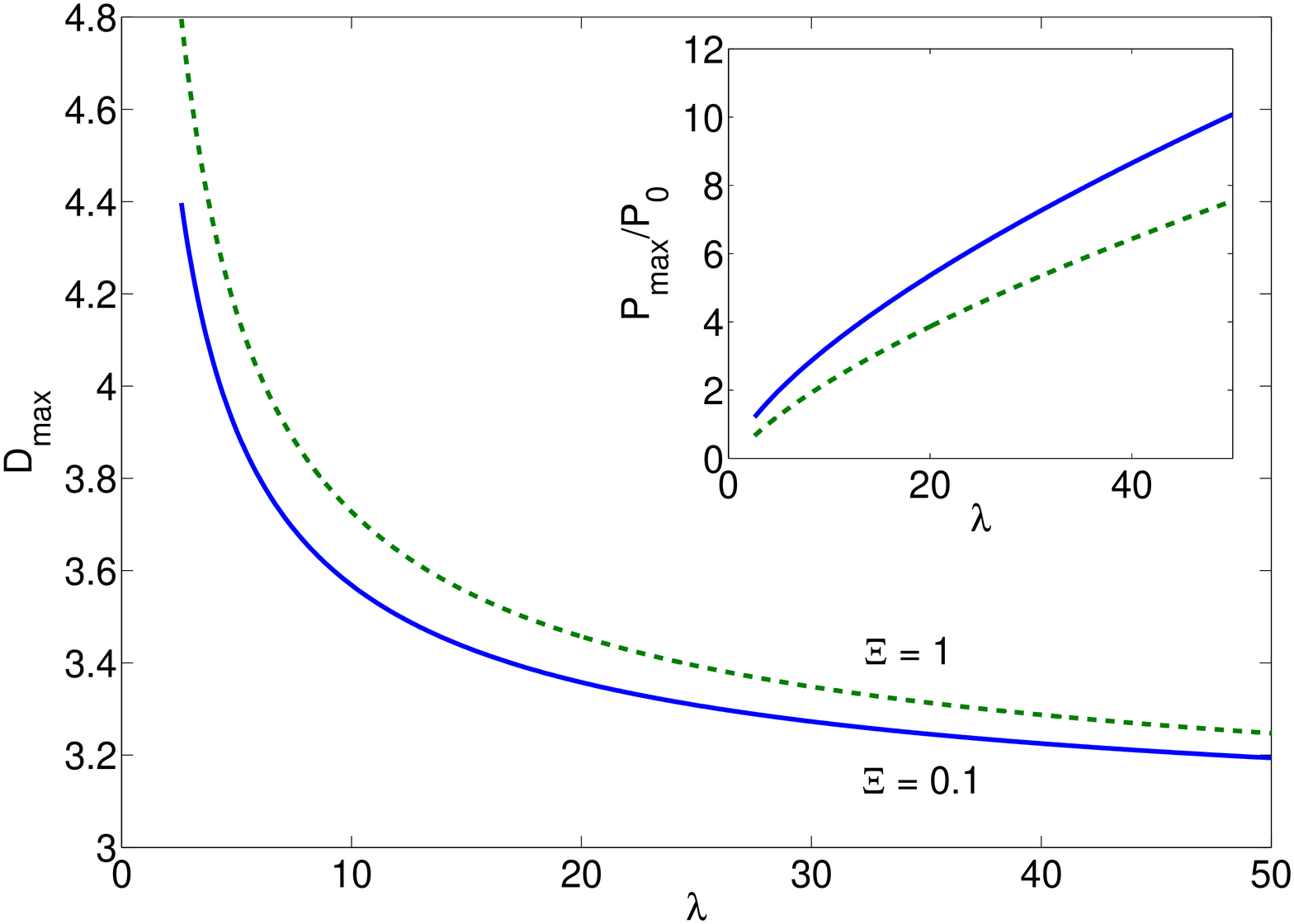}
\caption{The position and magnitude of the peak disjoining pressure as a function of fugacity for ions confined between like-charged surfaces. The main panel shows the surface separation at which the first peak in the disjoining pressure is observed, and shows that this moves away from $D=3$ as the fugacity decreases. The inset shows the concomitant decrease in the magnitude of the peak pressure.}
\label{pressure_fugacity}
\end{figure}

Figure \ref{pressure_fugacity} summarises the regime diagram for the system. Decreasing the fugacity shifts the first, most pronounced peak away from $D=3$, the minimum separation for which it is geometrically possible to fit 3 layers, and the peak pressure increases concomitantly. 
In the limit of large fugacity $\lambda \gg 1$, the peak disjoining pressure may be approximated by the asymptotic expressions
\begin{equation}
P_{\mathrm{max}} \approx \begin{cases} \frac{\lambda^2}{2^{1/3}},  &  \Xi \ll 1, \; \; \Xi \lambda \ll 1 \\ 
- \frac{\Xi}{4} & \Xi \gg 1. \end{cases}  
\end{equation} 
The position of the first peak in disjoining pressure can also be found asymptotically to be
\begin{equation}
D_{\mathrm{max}} \approx \begin{cases} 3 + \frac{48^{1/3}}{ \lambda^{2/3}},  &  \Xi \ll 1, \;\; \; \; \Xi \lambda \ll 1   \\ 
3 + \frac{2^{4/3}}{\lambda^{2/3}} & \Xi \gg 1. \end{cases}  
\label{D_max}
\end{equation} 

\section{Comparison with Experiment}

\begin{figure}
\includegraphics[scale=0.5]{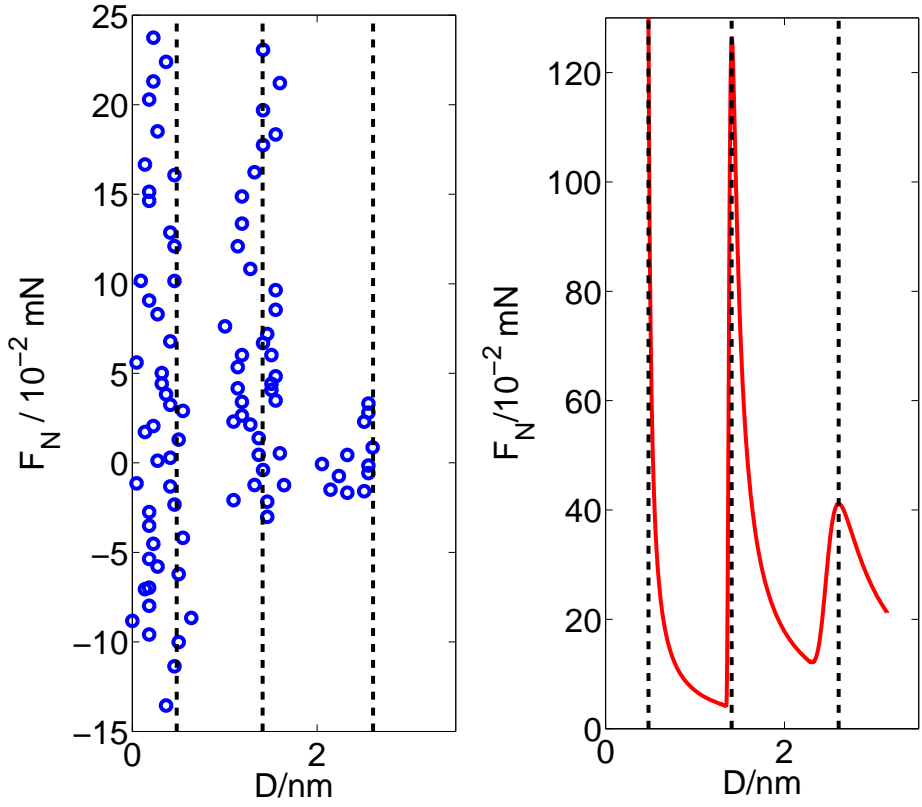}
\caption{Qualitative comparison of the normal force obtained in Ref \cite{perkin2011self} for the ionic liquid $\mathrm{[C_4C_1 Im][Tf_2 N]}$ confined between like-charged mica surfaces with the model prediction, the black dotted lines are guides to the eye. The parameters used are: $a=0.45 \AA$, $\lambda = 84 $ and $\Xi = 0.1$. We note that the microscopic force obtained by the model is scaled by $A_{g}/A_{i}$ to compare with the experimental data, which is actually preformed between crossed cylinders.}
\label{perkin_exp}
\end{figure}

Figure \ref{perkin_exp} compares experimental data from Ref \cite{perkin2011self} on the ionic liquid $\mathrm{[C_4C_1 Im][Tf_2 N]}$ with the predictions of the 1D model described here. The experiment is force controlled via a spring --- as such, the experiment registers branches with the gradient of the force larger than the spring constant as attractive ``jump in'' or repulsive ``jump-out''. This explained the qualitative difference between the theory and experiment.

We also want to estimate the typical size of the parameters $\Xi$ and $\lambda$. Although the cation and anion have different dimensions, an estimate based on the energy-optimised structure (the dimensions of the ions are obtained from Ref \cite{gebbie2013ionic}) suggests that $a_{\mathrm{cat}} = 0.35 \mathrm{nm} $ for the cation and $a_{\mathrm{ani}} = 0.55 \mathrm{nm} $ for the anion. Thus, as a rough guide, we take the mean and assume $a = 0.45 \mathrm{nm} $. The fugacity can be estimated from ionic liquid-air interfacial tension measurements. The chemical potential is the change in energy when removing a molecule from the bulk, thus it is twice the energy of transferring one molecule from the bulk to the interface where half of the interactions are deprived, \emph{viz.} $\mu \sim 2 \gamma \sigma_m$, where $\sigma_m$ is the area of a ``head group''. Experimentally, $\gamma  \approx 33.15 \; \text{mN/m}$ \cite{oliveira2012surface}, $\sigma_m \approx 27 \AA^2$ \cite{gebbie2013ionic}, thus $\mu = 4.4 k_B T$ and $\lambda \approx 84$. We note that this estimate is likely to be a lower bound. Chemically the ionic liquid ion is anisotropic, and preferred interactions can be maintained even when the ion is transferred from the bulk to the liquid-air interface, $e.g.$ by putting alkyl part outermost and charged group into the liquid. 

\begin{figure}
\includegraphics[scale=0.6]{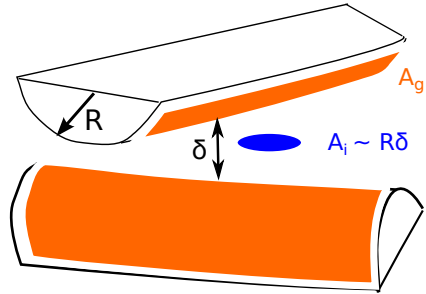}
\caption{A schematic illustration of the geometry of the crossed cylinder experimental apparatus used in \cite{perkin2011self}, showing the geometric area $A_g$ and interaction area  $A_{i}$}
\label{crossed_cylinder}
\end{figure}

Direct evaluation of the electrostatic parameter via Equation (\ref{eq_for_xi}) is difficult as $\sigma$ and $A$ are effective parameters that depend on the local arrangement of ions within a slab. However, the electrostatic parameter can be estimated with shear stress measurements. The yield force is the total force needed to ``unlock'' the slabs and slide them pass each other, whereas the static friction force gives an estimate of the contribution due to geometric incommensurability between the slabs and viscous dissipation. Therefore difference between the yield force and the static friction force, $\Delta F$, gives the typical force scale required to overcome the Coulomb attraction and slide the layers past each other for a typical slip distance $d_s$. Thus the typical macroscopic Coulomb interactions between the slabs is approximately $E \sim d_s \Delta F_s$. The corresponding microscopic interaction energy scale differs by a factor of $A_i/A_g$, where $A_{g}$ is the geometric area of the plates and $A_{i}$ is the interaction area, which is less than the geometric area due to the finite radius of curvature of the system (see Figure \ref{crossed_cylinder} for a schematic drawing). We use the estimate $A_{i} \approx R \delta$ where $R$ is the radius of curvature and $\delta$ is the typical surface separation, here taken as 1 \text{nm}. The typical slip length for $\mathrm{[C_4C_1 Im][Tf_2 N]}$ confined between mica surface is $d_s \sim 1 \mathrm{nm}$ and the typical force $\Delta F_s \sim 1 \mathrm{\mu N}$ \cite{smith2013quantized}, thus we have $\Xi \sim (d_s \Delta F_s/(k_B T)) (A_i/A_g) \sim 0.1$.   

We note that the theory shows only the electrostatic contribution to the normal force --- experimentally van der Waals interaction and the bulk pressure are important contributions. However, the only oscillatory component of the normal force is the electrostatic interaction, thus the qualitative alignment of the peaks predicted by the theory and observed in experiment is an important justification of the theory. In addition both van der Waals and bulk pressure are attractive forces, explaining the fact that the electrostatic component of the force is more positive (repulsive) than the experimentally measured force.

\section{Oppositely-Charged Surfaces} 
\begin{figure} 
\includegraphics[scale=0.45]{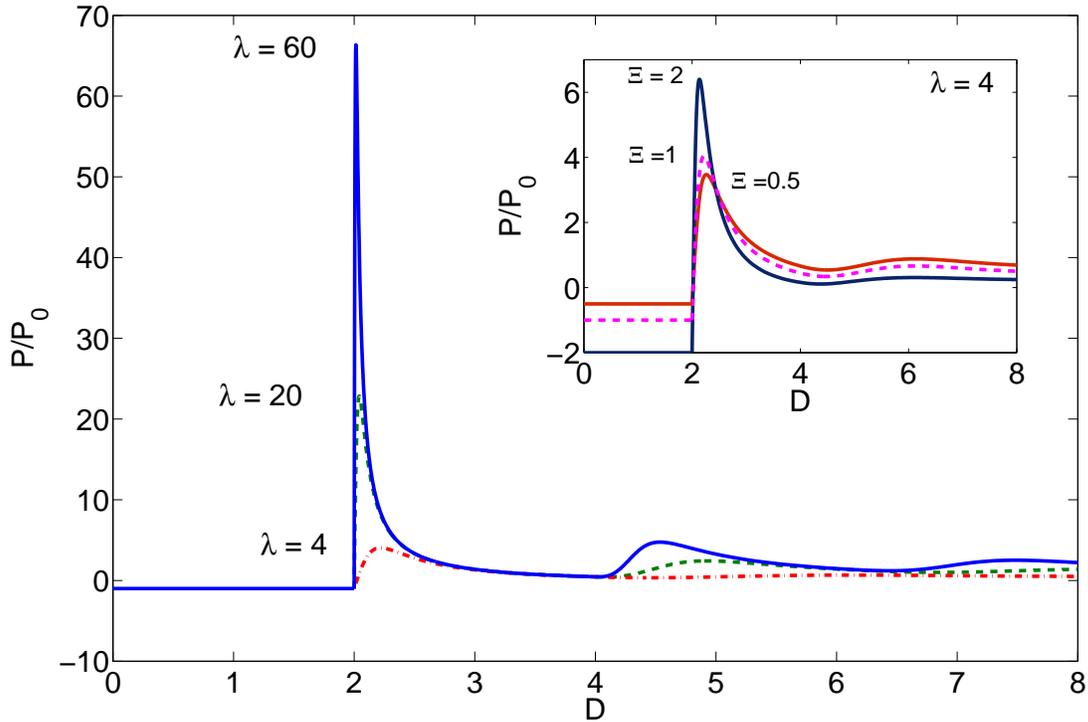}
\caption{The disjoining pressure as a function of separation for oppositely charged surface with $q=1$. The inset show the dependence of the pressure on the electrostatic parameter $\Xi$.}
\label{unlike_surface} 
\end{figure}

Similar considerations apply to ionic fluid confined between oppositely charged surfaces, with surface charge $\pm q \sigma$. The Coulomb gas model developed above can readily be applied. Figure \ref{unlike_surface} shows that the pressure is negative for small separations, corresponding to attractive interaction between the surfaces when the ions cannot effectively screen the surface charge. As in the case for like-charged surfaces, the adsorption transitions (and hence the maxima in disjoining pressure), become more pronounced as the fugacity of the ionic fluid increases. 

However, contrary to like-charged surfaces, the maxima in the disjoining pressure \emph{increases} as the electrostatic parameter $\Xi$ increases (see inset of Figure \ref{unlike_surface}). Consider for simplicity the case of $N=2$, in which the Hamiltonian for the ground state configuration of alternating positive and negative charges reads $H_2/(k_B T\; \Xi) = -D q^2 + y_2 (2q +1)$, where $y_2$ is the separation between the ion layers. As the Hamiltonian is linear in $y_2$, the extrema of the Hamiltonian are reached at the boundaries of allowed $y_2$, \emph{viz.} $y_2 = 1$ (corresponding to layers separated by the hard sphere diameter), and $y_2 = D-1$ (corresponding to layers located at the distance of closest approach to the charged surface). For $D>2$ the minimum $y_2$ is attained at $y_2 = D-1$, and ions are forced close to the surface, creating a ``cavity" at the centre which promotes abrupt adsorption of another layer of ions when the slit separation geometrically allows it. This is in direct contrast to the case of like-charged interfaces, where the combined electrostatic attraction of a layer to both surfaces pushes ions to the centre of the slit. 

\begin{figure} 
\includegraphics[scale=0.5]{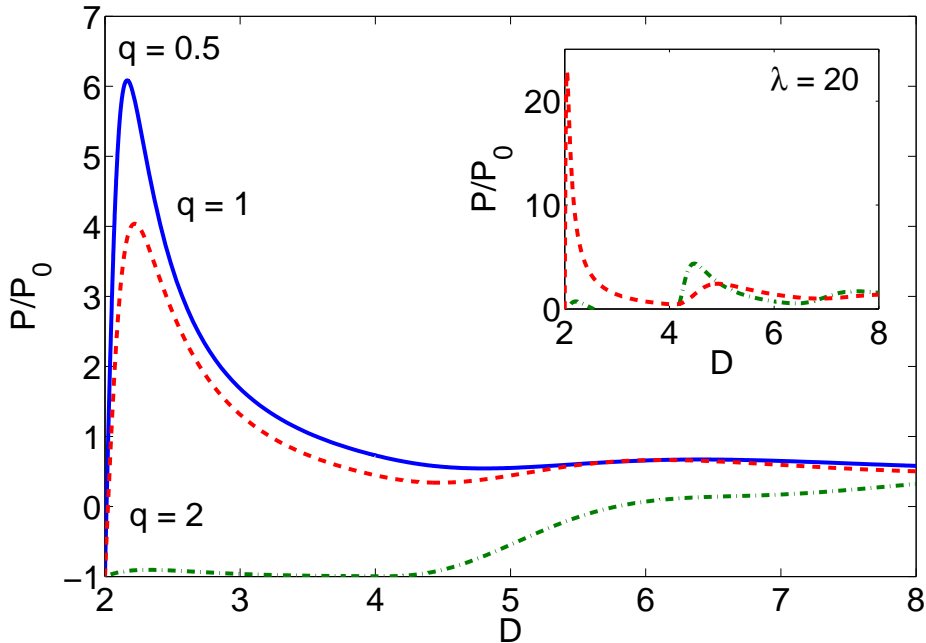}
\caption{The qualitative features of the disjoining pressure for ions confined between oppositely charged surfaces depends on whether the surfaces are overcharged ($q>1$) or undercharged ($q<1$). The main panel shows results with fugacity $\lambda=4$, and the inset shows $\lambda=20$.}
\label{unlike_surface_surcharge} 
\end{figure}

The qualitative behaviour of the disjoining pressure is also dependent on $q$, the ratio between the surface charge on the charged surface and the charge on the slabs. Figure \ref{unlike_surface_surcharge} show that the peak disjoining pressure is low for overcharged surface ($q>1$), as the strong electrostatic attraction between the surfaces is not effectively screened by the intervening ions, and the thermodynamic driving force for ion entry is overwhelmed by attraction between the surfaces. For overcharged surfaces the first peak in the disjoining pressure can be lower than subsequent peaks, whereby more ions have entered the system to screen the interaction between the surfaces.             

In the limit of small electrostatic parameter $\Xi$, the position of the first peak is given by 
\begin{equation}
D_{\mathrm{max}} = 2 + \frac{1}{\lambda} + \frac{\Xi}{6 \lambda} \left[3\lambda -2 + 3 q^2 (1-2 \lambda )\right] + O(\Xi^2).
\end{equation}
Thus increasing surface charge shifts the peak to larger separations, as the surface-surface attraction favours small surface separation (note that when $\lambda <1/2$, the first peak is located outside of the range $2<D<4$).  

We note that $q$ in this 1D model is an effective parameter --- physically it depends on the electrostatic interactions within a slab which is averaged out in this model. Experimentally for large monovalent ions and highly charged surfaces, surface charge cannot be completely neutralised even when the layers are laterally closed packed, thus $q>1$. On the other hand, for small ions or sparingly charged surfaces, ions can arrange themselves to form layers that completely neutralise or even overcompensate the surface charge, thus $q \le 1$. Therefore the qualitative form of the disjoining pressure is an effective way to interrogate not only the layered organisation of ions in the direction perpendicular to the surface, but also the lateral structure within the layer in the direction parallel to the surface. 

\section{Conclusion}
We have considered the equilibrium properties of an ionic liquid confined between charged surfaces using a 1D Coulomb gas model. The theory reveals the roles of all physical parameters, but most importantly they group themselves into two dimensionless parameters: the fugacity of the bulk fluid $\lambda$ and the electrostatic parameter $\Xi$. These two parameters have clear constituents, and can be independently measured or estimated. 

The model shows that the disjoining pressure decays in an oscillatory manner with the separation between surfaces, with the maxima corresponding to entering of discrete ``layers'' of ions, and the peaks in disjoining pressure becoming more pronounced as the fugacity increases. For like-charged surfaces, the peak disjoining pressure decreases with increasing electrostatic parameter as the ground state Hamiltonian is independent of the number of layers in-between the surfaces. The theory is in good qualitative agreement with experimental data on $\mathrm{[C_4C_1 Im][Tf_2 N]}$ confined between mica surfaces, with all parameters independently estimated without fitting.  

For oppositely charged surfaces, the theory predicts that increasing the electrostatic parameter increases the peak disjoining pressure. Ions are pulled close to the charged surfaces, creating a ``cavity'' at the centre of the slit that allows adsorption of ions from the bulk. The electrostatic parameter can be varied by altering the effective ion radius and the charge on the ionic liquid ions. Experimental studies with atomically flat, oppositely charged surfaces are currently scarce. We hope that our model will motivate further experimental and computational studies. 


\begin{acknowledgments}
We thank J S Wettlaufer for insightful discussions. This work is supported by an EPSRC Research Studentship to AAL.  
\end{acknowledgments}

\bibliography{ref_nanofilm}
\end{document}